\documentstyle[preprint,aps,tighten]{revtex}
\begin{document}

\title{ Ab-initio calculations of exchange interactions, spin-wave
stiffness constants, and Curie temperatures of Fe, Co, and Ni}

\author{M.~Pajda$^{\; a}$, J.~Kudrnovsk\'y$^{\; b,a}$, I.~Turek$^{\; c,d}$,
V.~Drchal$^{\;b}$, and P.~Bruno$^{\; a}$
\\
$^a$ Max-Planck-Institut f\"ur Mikrostrukturphysik, Weinberg 2,
     D-06120 Halle, Germany
\\
$^b$ Institute of Physics, Academy of Sciences of the Czech Republic,
     Na Slovance 2,\\ CZ-182 21 Prague 8, Czech Republic
\\
$^c$ Institute of Physics of Materials, Academy of Sciences of the
     Czech Republic, \v{Z}i\v{z}kova 22, CZ-616 62 Brno,
     Czech Republic
\\
$^d$ Department of Electronic Structures, Charles University,\\
Ke Karlovu 5, CZ-121 16 Prague 2, Czech Republic }

\maketitle
\nopagebreak
\begin{abstract}

We have calculated Heisenberg exchange parameters for bcc-Fe,
fcc-Co, and fcc-Ni using the non-relativistic spin-polarized
Green function technique within the tight-binding linear
muffin-tin orbital method and by employing the magnetic force
theorem to calculate total energy changes associated with a local
rotation of magnetization directions. We have also determined
spin-wave stiffness constants and found the dispersion curves for
metals in question employing the Fourier transform of calculated
Heisenberg exchange parameters. Detailed analysis of convergence
properties of the underlying lattice sums was carried out and a
regularization procedure for calculation of the spin-wave
stiffness constant was suggested. Curie temperatures were
calculated both in the mean-field approximation and within the
Green function random phase approximation. The latter results
were found to be in a better agreement with available
experimental data.
\end{abstract}


\section{Introduction}
\label{sec_Intr}

The quantitative description of thermodynamic properties of magnetic metals
is challenging solid state theorists since decades. Thanks to the development
of density functional theory and its implementation into {\em ab
initio\/} computational schemes, an excellent understanding of their ground
state (i.e., at $T=0$~K) has been achieved. On the other hand, most of the
progress towards a description of magnetic metals at non-zero temperature has
been based upon models in which the electronic structure is oversimplified
and described in terms of empirical parameters. Although this
approach has the great merit of emphasizing the
relevant mechanisms and concepts, it cannot properly take into account
the complex details of the electronic structure and is therefore unable to
yield quantitative predictions of the relevant physical quantities such as
spin-wave stiffness, Curie temperature $T_C$, etc., for comparison with experimental
data.

It is therefore of a great importance to develop an {\em ab initio\/},
parameter-free, scheme for the description of ferromagnetic metals
at $T > 0$~K. Such an
approach must be able to go beyond the ground state and to take into
account excited states, in particular the magnetic excitations responsible
for the decrease of the magnetization with temperature and for the phase
transition at $T=T_C$. Although density functional theory can be formally
extended to non-zero temperature, there exists at present no practical
scheme allowing to implement it. One therefore has to rely on approximate
approaches. The approximations to be performed must be chosen on the basis of
physical arguments.

In itinerant ferromagnets, it is well known that magnetic excitations are
basically of two different types:
(i) Stoner excitations, in which an electron
is excited from an occupied state of the majority-spin band to an empty state
of the minority-spin band and
creates an electron-hole pair of triplet spin. They are associated with
longitudinal fluctuations of the magnetization;
(ii) the spin-waves or
magnons, which correspond to collective transverse fluctuations of the
direction of
the magnetization.
Near the bottom of the excitation spectrum, the density of
states of magnons is considerably larger than that of corresponding Stoner
excitations, so that the thermodynamics in the low-temperature regime is
completely dominated by magnons and Stoner excitations can be neglected.
Therefore it seems reasonable
to extend this approximation up to the Curie temperature, and to estimate the
latter by neglecting Stoner excitations. This is a good approximation
for ferromagnets with a large exchange splitting such as Fe and
Co, but it is less justified for Ni which has a small exchange splitting.

The purpose of the present paper is to describe the spin-wave properties
of transition metal itinerant ferromagnets at {\em ab initio\/} level.
With thermodynamic properties in mind, we are primarily interested in the
long-wavelength magnons with the lowest energy. We shall adopt the
{\em adiabatic approximation\/} in which the {\em precession\/}
of the magnetization due to a spin-wave is neglected when calculating the
associated change
of electronic energy. Clearly, the condition of validity of this approximation
is that the precession time of the magnetization should be large as
compared to characteristic times of electronic motion, namely, the hopping
time of an electron from a given site to a neighboring one, and the precession
time of the spin of an electron subject to the exchange field. In other words,
the spin-wave energies should be small as compared to the band width and to
the exchange splitting. This approximation becomes exact in the limit of long
wavelength magnons, so that the spin-wave stiffnesses constant calculated in this way
are in principle exact.

This procedure corresponds to a mapping of the itinerant electron system onto
an effective Heisenberg Hamiltonian with classical spins
\begin{equation}
H_{\mbox{\scriptsize eff}} = - \sum_{i\neq j}
J_{ij} {\bf e}_i \, \cdot {\bf e}_j \, ,
\label{eq_HH}
\end{equation}
where $J_{ij}$ is the exchange interaction energy between two particular
sites $(i,j)$, and ${\bf e}_i, {\bf e}_j$ are unit vectors pointing in the
direction of local magnetic moments at sites $(i,j)$, respectively.
The same point of view
has been adopted in various papers recently published on the same topic
\cite{ Liechtenstein1984, Liech, Antropov1996, Halilov1998, Antropov1999,
Brown1999, Schilfgaarde1999, Ivanov1999, MacLaren1999, Ujfalussy1999,
Sonia, Sakuma, Spis, Antr2}.

The procedure for performing the above mapping onto an effective
Heisenberg Hamiltonian relies on the constrained density functional theory
\cite{ Dederichs1984}, which allows to obtain the ground state energy for
a system subject to certain constraints. In the case of magnetic interactions,
the constraint consists in imposing a given configuration of spin-polarization
directions, namely, along ${\bf e}_i$ within the atomic cell $i$. Note that
{\em intracell\/} non-collinearity of the spin-polarization is neglected
since we are primarily interested in low-energy excitations due to
{\em intercell\/} non-collinearity.

Once the exchange parameters $J_{ij}$ are obtained, the spin-dynamics
\cite{ Halilov1998, Niu1998, Niu1999} can be determined from the effective
Hamiltonian (\ref{eq_HH}) and one obtains the result known from spin-wave
theories of localized ferromagnets: the spin-wave energy $E({\bf q})$ is
related to the exchange parameters $J_{ij}$ by a simple Fourier
transformation
\begin{equation}
E({\bf q})=
\frac{4\mu_B}{M}\sum_{j\neq 0}J_{0j}(1-\exp(i{\bf q} \cdot {{\bf R}_{0j}})) \, ,
\label{eq_E}
\end{equation}
where ${\bf R}_{0j}$=${\bf R}_{0}-{\bf R}_{j}$ denote lattice vectors in
the real space, ${\bf q}$ is a vector in the corresponding Brillouin
zone, and $M$ is the magnetic moment per atom ($\mu_B$ is the Bohr
magneton).

There are basically two approaches to calculate the exchange
parameters and spin-wave energies. The first one which we adopt
in the present paper, referred to as the real-space approach,
consists in calculating directly $J_{ij}$ by employing the change
of energy associated with a constrained rotation of the
spin-polarization axes in cells $i$ and $j$ \cite{ Liech}. In the
framework of the so-called magnetic force-theorem \cite{ Liech,
Dede} the change of the total energy of the system can be
approximated by the corresponding change of one-particle energies
which significantly simplifies calculations. The spin-wave
energies are then obtained from Eq.~(\ref{eq_E}). In the second
approach, referred to as the frozen magnon approach, one chooses
the constrained
 spin-polarization configuration to be the one of
a spin-wave with the wave vector ${\bf q}$ and computes $E({\bf q})$
directly by employing the generalized Bloch theorem for a spin-spiral
configuration \cite{ Sandratskii1991}. Like the above one,
approach can be implemented with or without using the the
magnetic force-theorem.
Both the real-space approach and the frozen magnon approach
can be implemented by using either a finite or an infinitesimal
rotation, the latter choice is usually preferable.
The exchange parameters $J_{ij}$ are then obtained by inverting
Eq.~(\ref{eq_E}). One should also mention a first-principles theory
of spin-fluctuations (the so-called disordered local moment picture)
based on the idea of a generalized Onsager cavity field \cite{ Stau}.

The spin-wave stiffness $D$ is given by the curvature of the spin-wave
dispersion $E({\bf q})$ at ${\bf q}=0$.
Although its calculation is {\em in principle\/} straightforward in the
real-space approach, we shall show that serious difficulties arise due
to the Ruderman-Kittel-Kasuya-Yoshida (RKKY) character of magnetic
interactions in metallic systems.
These difficulties have been underestimated in a number of previous
studies \cite{ Liech, Antropov1999, Spis, Antr2},
and the claimed agreement with experiment is thus fortituous. We shall
present a procedure allowing to overcome these difficulties.
In addition, we shall demonstrate that the evaluation of the spin-wave
dispersion $E({\bf q})$ in the real-space approach has to be also done
carefully with respect to the convergency of results with the number of
shells included.

Finally, to obtain thermodynamic quantities such as the Curie
temperature, we apply statistical mechanics to the effective
Hamiltonian (\ref{eq_HH}). In the present paper, we use two
different approaches to compute the Curie temperature. The first
one is the commonly used mean field approximation (MFA). The
limitations of this method are well known: it is correct only in
the limit of high temperatures (above $T_C$), and it fails to
describe the low-temperature collective excitations (spin-waves).
The second approach is the Green function method within the
random phase approximation (RPA) \cite{ Tyab, Tahi, Tahi2, Call,
Tahi3, Wang}. The RPA is valid not only for high temperatures,
but also at low temperatures, and it describes correctly the
collective excitations (spin-waves). In the intermediate regime
(around $T_C$), it represents a rather good approximation which
may be viewed as an interpolation between the high and low
temperature regimes. It usually yields a better estimate of the
Curie temperature as compared to the MFA. It should be noted,
however, that both the MFA and RPA fail to describe correctly the
critical behavior and yield in particular incorrect critical
exponents.

\section{Formalism}
\label{sec_Form}

The site-off diagonal exchange interactions $J_{ij}$ are calculated using
the expression \cite{ Liech}
\begin{equation}
J_{ij} =\frac{1}{4\pi} \, {\rm Im} \int_{C} \, {\rm tr}_L \,
\left[ (P_{i}^{\uparrow}(z)-P_{i}^{\downarrow}(z)) \,
g^{\uparrow}_{ij}(z) \, (P_{j}^{\uparrow}(z)-P_{j}^{\downarrow}(z)) \,
g^{\downarrow}_{ji}(z) \right] \, {\rm d} z \, ,
\label{eq_Jij}
\end{equation}
which is evaluated in the framework of the first-principles tight-binding
linear muffin-tin orbital method (TB-LMTO) \cite{ OKA}.
Here ${\rm tr}_L$ denotes the trace over the angular momentum $L=(\ell m)$,
energy integration is performed in the upper half of the complex energy
plane over a contour $C$ starting below the bottom of the valence band
and ending at the Fermi energy,
$P^{\sigma}_i(z)$ are diagonal matrices of the so-called potential
functions of the TB-LMTO method for a given spin direction
$\sigma=\uparrow,\downarrow$ with elements $P^{\sigma}_{i,L}(z)$ ,
and $g^{\sigma}_{ij}(z)$ are the so-called auxiliary Green
function matrices with elements $g^{\sigma}_{iL,jL'}(z)$ \cite{ Turk}
defined as
\begin{equation}
[g^{\sigma}(z)]_{iL,jL'}^{-1}=P_{i,L}^{\sigma}(z) \, \delta_{LL'}
\, \delta_{i,j} - S_{iL,jL'} \, .
\label{eq_GF}
\end{equation}
We have also introduced the spin-independent screened structure constant
matrix
$S_{i,j}$ with elements $S_{iL,jL'}$ which characterizes the underlying
lattice within the TB-LMTO approach \cite{ Turk}.

Calculated exchange parameters were further employed to estimate the
spin-wave
spectrum $E({\bf q})$ as given by Eq.~(\ref{eq_E}).
For cubic systems and in the range of small ${\bf q}$ we have
\begin{equation}
E({\bf q})= D \, {q}^2 \, ,
\label{eq_LD}
\end{equation}
where $q=|\bf q|$.
The spin-wave stiffness coefficient $D$ can be expressed
directly in terms of the
exchange parameters $J_{0j}$ as \cite{ Liech}
\begin{equation}
D=\frac{2\mu_B}{3M}
\sum_{j} \, J_{0j} \, R_{0j}^2 \, ,
\label{eq_D}
\end{equation}
where $R_{0j}=|{\bf R}_{0j}|$.
The summation in Eq.~(\ref{eq_D}) runs over all sites but in practice
the above sum has to be terminated at some maximal value of
$R_{0j}=R_{max}$.
There is a lot of misunderstanding in the literature as concerns
the use of Eq.~(\ref{eq_D}).
Several calculations were done with $R_{max}$ corresponding to
the first few coordination shells \cite{ Liech, Spis, Antr2}.
In other calculations \cite{ Antr2, Antropov1999} the authors realized
the problem of the termination of $R_{max}$ but they did not suggest
an appropriate method to perform the sum ((\ref{eq_D}) in the direct
space.
We will demonstrate that terminating the sum in Eq.~(\ref{eq_D}) after
some value of $R_{max}$ is fundamentally incorrect because it
represents a non-converging quantity and we will show how to resolve
this problem from a numerical point of view.
The reason for such behavior is the long-range oscillatory character of
$J_{ij}$ with the distance $R_{ij}$ in ferromagnetic metals.

Alternatively, it is possible to evaluate $E({\bf q})$ directly
in the reciprocal space \cite{ Antropov1999} as
\begin{eqnarray}\label{eq_Jqd}
E({\bf q}) &=& \frac{4\mu_B}{M} \, (J({\bf 0})-J({\bf q})) \, ,
\\ \nonumber
J({\bf q}) &=& \frac{1}{4\pi N} \,{\rm Im} \, \sum_{\bf k} \, \int_{C}
\, {\rm tr}_L \, \left[ (P^{\uparrow}(z)-P^{\downarrow}(z)) \,
g^{\uparrow}({\bf k+q}, z) \, (P^{\uparrow}(z)-P^{\downarrow}(z)) \,
g^{\downarrow}({\bf k}, z) \right] \, {\rm d} z \, ,
\end{eqnarray}
and to determine the spin-stiffness constant as a second derivative
of $E({\bf q})$ with respect to ${\bf q}$.


Calculated exchange parameters can be also used to determine Curie
temperatures of considered metals.
Within the MFA
\begin{equation}\label{eq_TcMFA}
k_B T_C^{MFA} =\frac{2}{3}\sum_ {j \ne 0}J_{0j}=\frac{M}{6 \mu_B} \,
\frac{1}{N}\sum_{\bf q} \, E({\bf q}) \, ,
\label{eq_MFA}
\end{equation}
where $E({\bf q})$ is the spin-wave energy (\ref{eq_E}).
We have calculated $T_C^{MFA}$ directly from the expression
$k_B T_C^{MFA}=2 J_{0}/3$, where \cite{ Liech}
\begin{eqnarray}\label{eq_sumrule}
J_{0} \equiv \sum_{i \ne 0} J_{0i} = -\ \frac{1}{4\pi}\int_{C} \,
{\rm Im \, tr}_L \,
\left[ (P_{0}^{\uparrow}(z)-P_{0}^{\downarrow}(z)) \,
(g^{\uparrow}_{00}(z)-g^{\downarrow}_{00}(z)) \, +  \right.
\nonumber \\
\left. (P_{0}^{\uparrow}(z)-P_{0}^{\downarrow}(z)) \,
g^{\uparrow}_{00}(z) \, (P_{0}^{\uparrow}(z)-P_{0}^{\downarrow}(z)) \,
g^{\downarrow}_{00}(z) \right] \, {\rm d} z \, .
\label{eq_J0}
\end{eqnarray}
The expression for the Curie temperature within the GF-RPA approach is
\cite{ Wang}
\begin{equation}\label{eq_TcRPA}
\frac{1}{k_B T_C^{RPA}}=\frac{6 \mu_B}{M} \,
\frac{1}{N}\sum_{\bf q} \frac{1}{E(\bf q)} \, .
\label{eq_RPA}
\end{equation}
The integrand in (\ref{eq_RPA}) is singular for ${\bf q}=0$.
We have therefore calculated $T_C^{RPA}$ using the expression
\begin{eqnarray}
\frac{1}{k_B T_C^{RPA}} &=& -\lim_{z \to 0} \, \frac{6 \mu_B}{M} \;
{\rm Re} \, G_{m}(z) \, ,
\nonumber \\
G_{m}(z) &=& \frac{1}{N} \sum_{\bf q} {\frac{1}{z-E({\bf q})}} \, .
\label{eq_RPA1}
\end{eqnarray}
The quantity $G_{m}(z)$ is the magnon Green function corresponding to
the dispersion law $E({\bf q})$ and it was evaluated for energies $z$ in
the complex energy plane and its value for $z=0$ was obtained using the
analytical deconvolution technique \cite{ deconv}.
It should be noted that the MFA and the RPA differ essentially in the
way in which they weight various $J_{ij}$, namely more distant neighbors
play a more important role in the RPA as compared to the MFA.
It is seen from Eqs.~(\ref{eq_MFA},\ref{eq_RPA}) that $T_C^{MFA}$
and $T_C^{RPA}$ are given as the arithmetic and harmonic averages of
the spin-wave energies $E({\bf q}) $, respectively, and therefore
it holds $T_C^{MFA} > T_C^{RPA}$.

\section{Results and discussion}
\subsection{Details of calculations}

Potential functions and Green functions which appear in
Eq.~(\ref{eq_Jij}) were determined within the non-relativistic
TB-LMTO method in the so-called orthogonal representation
\cite{ Turk} assuming the experimental lattice
constants and the exchange potential in the form suggested by
Vosko-Wilk-Nusair \cite{ Vosk}. It should be noted that some
calculations, in particular for $T_{C}^{MFA}$, were also done
using the scalar-relativistic formulation. The contour integral
along the path $C$ which starts below the lowest occupied band
and ends at the Fermi energy (we assume zero temperature) was
calculated following the scheme described in \cite{ Turk} which
employs the Gaussian quadrature method. Twenty energy nodes were
used on the semi-circle in the upper part of the complex energy
plane. The integration over the full Brillouin zone was performed
very carefully to obtain well-converged results even for very
distant coordination shells (up to 172-nd shell for fcc lattice
and the 195-th shell for bcc lattice). In particular, we have
used up to $5 \times 10^6$ ${\bf k}$-points in the full Brillouin
zone for the energy point on the contour $C$ closest to the Fermi
energy, and the number of ${\bf k}$-points then progressively
decreased for more distant points, and for points close to the
bottom of the band.

\subsection{Effective Heisenberg exchange parameters} \label{efhh}

We will first discuss qualitatively the dependence of $J_{ij}$ on the
distance $R_{ij}=|{\bf R_i - R_j}|$.
In the limit of large values of $R_{ij}$ the expression (\ref{eq_Jij})
can be evaluated analytically by means of the stationary phase
approximation \cite{ PBr}.
For simplicity we consider here a single-band model but the results
can be generalized also to the multiband case (see
Ref.~\onlinecite{ IEC}).
For a large $R_{ij}$ behaves $g^{\sigma}_{ij}$ as
\begin{equation}
g_{ij}^{\sigma}(E + i 0^{+}) \propto
\frac{\exp \left[ i ({\bf k}^{\sigma} \cdot {\bf R}_{ij} + \Phi^{\sigma})
\right]} {R_{ij}} \, ,
\end{equation}
where ${\bf k}^{\sigma}$ is the wave vector of energy $E$ in a direction
such that the associated group velocity $\nabla_{\bf k} E^{\sigma}({\bf k})$
is parallel to ${\bf R}_{ij}$, and $\Phi^{\sigma}$ denotes a corresponding
phase factor.
The energy integration  in (\ref{eq_Jij}) yields additional factor of
$1/R_{ij}$ \cite{ PBr} and one obtains
\begin{equation}
J_{ij} \propto  {\rm Im} \;
\frac{ \exp \left[ i ({\bf k}_{\rm F}^{\uparrow} +
{\bf k} _{\rm F}^{\downarrow}) \cdot {\bf R}_{ij} + \Phi^{\uparrow} +
\Phi^{\downarrow} \right] } {R_{ij}^{3}} \, .
\label{eq_Jwf0}
\end{equation}
For a weak ferromagnet both Fermi wave vectors ${\bf k}_{\rm F}^{\uparrow}$
and ${\bf k}_{\rm F}^{\downarrow}$ are real and one obtains a characteristic
RKKY-like behavior
\begin{equation}
J_{ij} \propto
\frac{ \sin \left[ ({\bf k}_{\rm F}^{\uparrow} +
{\bf k} _{\rm F}^{\downarrow}) \cdot {\bf R}_{ij} + \Phi^{\uparrow} +
\Phi^{\downarrow} \right] } {R_{ij}^{3}} \, ,
\label{eq_Jwf}
\end{equation}
i.e., the exchange interaction has an oscillatory character with an
envelope decaying as $1/R_{ij}^{3}$.
On the other hand, for a strong ferromagnet with a fully occupied majority
band the corresponding Fermi wave vector is imaginary, namely
${\bf k}_{\rm F}^{\uparrow}=i \mbox{\boldmath$ \kappa$}_{\rm F}^{\uparrow}$
and one obtains an exponentially  damped RKKY behavior
\begin{equation}
J_{ij} \propto \frac{ \sin ({\bf k}_{\rm F}^{\downarrow} \cdot {\bf R}_{ij}
+ \Phi^{\uparrow} + \Phi^{\downarrow}) \,
\exp ( -\ \mbox{\boldmath$ \kappa$}_{\rm F}^{\uparrow} \cdot
{\bf R}_{ij}) } {R_{ij}^{3}} \, .
\label{eq_Jsf}
\end{equation}
The qualitative features of these RKKY-type oscillations of $J_{ij}$
will not be changed in realistic ferromagnets.
For a weak ferromagnet, like Fe, one expects a pronounced RKKY character
giving rise to strong Kohn anomalies in the spin-wave spectrum.
On the other hand, for Co and Ni which are almost strong ferromagnets
one expects a less pronounced RKKY character, less visible Kohn anomalies
in the spin-wave spectrum (see Sec.~\ref{sec_disp}), and faster decay of
$J_{ij}$ with a distance $R_{ij}$.
It should be noted that due to the {\it sp-d} hybridization no itinerant
ferromagnet is a truly strong ferromagnet.

The calculated Heisenberg exchange parameters $J_{ij}$ for
bcc-Fe, fcc-Co, and fcc-Ni are presented in the Table I for the
first ten shells. The exchange parameters $J_{ij}$ remain
non-negligible over a very long range along the [111]-direction,
and change from ferromagnetic to antiferromagnetic couplings
already for the third nearest-neighbors (NN). In case of Co this
change appears only for the 4th NN whereas Ni remains
ferromagnetic up to the 5th NN. It should be noted a short range
of $J_{ij}$ for the case of Ni, being essentially a decreasing
function of the distance with the exception of the second NN.
Such behavior is in a qualitative agreement with conclusions
obtained from the asymptotic behavior of $J_{ij}$ with distance
discussed above, in particular with the fact that bcc-Fe is a
weak ferromagnet while fcc-Co and, in particular, fcc-Ni are
almost strong ferromagnets. There have been several previous
calculations of $J_{ij}$'s for Fe and Ni \cite{ Liech,
Antropov1999,  Schilfgaarde1999, Sonia, Wang}. Present
calculations agree well with calculations of Refs.~\cite{ Liech,
Antropov1999, Sonia} and there is also a reasonable agreement
with results of Refs.~\cite{ Wang, Schilfgaarde1999}. It should
be mentioned that $J_{ij}$ for both fcc-Co and hcp-Co were
determined and they agree quite well with each other \cite{
Schilfgaarde1999} (see also Table III below).
Finally, we have also verified numerically the validity of
important sum rule, namely $J_{0}=\sum_{i \ne 0} J_{0i}$.
The sum fluctuates with the number of shells very weakly for,
say, more than 50 shells.

\subsection{Dispersion relations} \label{sec_disp}
Calculated magnons energy spectra  $E({\bf q})$ along the high symmetry
directions of the Brillouin zone are presented in Figs.~\ref{Fig.1}~a--c
together with available experimental data
\cite{ Loong1984, Mook1985, Lynn1975}.
We have used all calculated shells to determine $E({\bf q})$, namely 195
and 172 shells for bcc- and fcc-metals, respectively.
Corresponding plot of $E({\bf q})$ for fcc-Ni exhibits parabolic, almost
isotropic behavior for long wavelengths and a similar behavior is also
found for fcc-Co.
On the contrary, in bcc-Fe we observe some anisotropy of $E({\bf q})$,
i.e., $E({\bf q})$ increases faster along the $\Gamma-N$ direction
and more slowly along the $\Gamma-P$ direction.
In agreement with Refs.~\onlinecite{ Halilov1998, Schilfgaarde1999}
we observe a local minima around the point $H$ along $\Gamma -H$ and
$H-N$ directions in the range of short wavelengths.
They are indications of the so-called Kohn anomalies \cite{ Halilov1998}
which are due to long-range interactions mediated by the RKKY interactions
similarly like Kohn-Migdal anomalies in phonon spectra are due to
long-range interactions mediated by Friedel oscillations.
It should be mentioned that minima in dispersion curve of bcc-Fe appear
only if the summation in (\ref{eq_E}) is done over a sufficiently large
number of shells, in the present case for more than 45 shells.
A similar observation concerning of the spin-wave spectra of bcc-Fe was
also done by Wang et al. \cite{ Wang} where authors used the fluctuating
band theory method using semiempirical approach based on a fitting
procedure for parameters of the Hamiltonian.
On the other hand in a recent paper by Brown et al. \cite{ Brown1999}
above-mentioned Kohn-anomalies in the behavior of spin-wave spectra of
bcc-Fe were not found, possibly because the spin-wave dispersion was
obtained as an average over all directions in the ${\bf q}$-space.

Present results for dispersion relations compare well with
available experimental data of measured spin-wave spectra for Fe
and Ni \cite{ Loong1984, Mook1985, Lynn1975}. For low-lying part
of spectra there is also a good agreement of present results for
dispersion relations with those of Refs.~\cite{ Halilov1998,
Schilfgaarde1999} obtained using the frozen magnon approach.
There are, however, differences for a higher part of spectra, in
particular for the magnon bandwidth of bcc-Fe which can be
identified with the value of $E({\bf q})$ evaluated at the
high-symmetry point ${\bf q}={\rm H}$ in the bcc-Brillouin zone.
The origin of this disagreement is unclear.
We have carefully checked the convergence of the magnon dispersion
laws $E({\bf q})$ with the number of shells included in
Eq.~(\ref{eq_E}) and it was found to be weak for 50 -- 70 shells
and more.
However, if the number of shells is small the differences may be
pronounced, e.g., our scalar-relativistic calculations give for
the bcc-Fe magnon bandwiths the values of 441 meV and 550 meV for
15 and 172 shells, respectively.
The former value agrees incidentally very well with that given
in Refs.~\cite{ Halilov1998, Schilfgaarde1999}.
On the other hand, even small differences in values of $E({\bf q})$
are strongly amplified when one evaluates the second derivative of
$E({\bf q})$ with respect to ${\bf q}$, i.e., the spin-wave
stiffness constant.
One should keep in
mind, however, that the above discussion is somehow academic, for
it concerns an energy region where the adiabatic approximation
ceases to be a good one, so that spin-waves are non longer well
defined because of their strong damping due to Stoner excitations
(see e.g. \cite{ Antropov1999}.
The results of theoretical calculations based upon the adiabatic
approximation can be thus compared with each other, but not
with experimental data.
It should be pointed out that the influence of deviations in
the calculation of magnon spectra for large values of ${\bf q}$
of the Curie temperature is minimized for its RPA value as
compared to its MFA value (see Eqs.~(\ref{eq_TcMFA},\ref{eq_TcRPA}).

\subsection{Spin-wave stiffness constant} \label{sec_sws}

As was already mentioned the sum in (\ref{eq_D}) does not converge
due to the characteristic RKKY behavior (\ref{eq_Jwf}) and, therefore,
Eq.~(\ref{eq_D}) cannot be used directly to obtain reliable values for
the spin-wave stiffness constant.
This is demonstrated in Fig.~\ref{Fig.3} where the dependence of
calculated spin-wave stiffness constants on the parameter $R_{max}$
in Eq.~(\ref{eq_D}) is plotted.
The oscillatory character of $D$ versus $R_{max}$ persists for large
values of $R_{max}$ for the case of bcc-Fe and even negative values of
spin-wave stiffness constants were obtained for some values of $R_{max}$.
To resolve this difficulty we suggest to regularize the expression
(\ref{eq_D}) by substituting it  by the formally equivalent
expression which is, however, numerically convergent
\begin{eqnarray}
D&=&\lim_{\eta \to 0} \, D(\eta) \, ,
\nonumber \\
D(\eta) &=& \lim_{R_{max} \to \infty} \frac{2\mu_B}{3M}
\sum_{0<R_{0j}\le{R_{max}}} \, J_{0j} \, R_{0j}^2 \,
\exp(-\eta R_{0j}/a) \, .
\label{eq_Dc}
\end{eqnarray}
The quantity $\eta$ plays a role of a damping parameter which
makes the sum over $R_{ij}$ absolutely convergent as it is seen
from Fig.~\ref{Fig.4}. The quantity $ D(\eta)$ is thus an
analytical function of the variable $\eta$ for any value $\eta >
0$ and can be extrapolated to the value $\eta=0$. To show that
the limit for $\eta \to 0$ is indeed finite and that our scheme
is mathematically sound, let us consider as an example a typical
RKKY interaction $J(R)\propto \sin(kR + \Phi) / R^3 $. For large
R we can employ Eq.~(\ref{eq_Jwf}) and substitute the sum in
(\ref{eq_D}) by a corresponding integral. We obtain
\begin{equation}
\lim_{\eta \to 0} \, D(\eta) \propto 4 \pi \int_{R_{0}}^{\infty}
\, R^{2} \, \frac{{\sin(kR+\Phi})}{R^{3}} \, {\rm d} R =
- \cos(\Phi) \, {\rm si}(kR_{0}) - \sin(\Phi) \, {\rm ci}(kR_{0})
\, ,
\end{equation}
where si and ci are integral sine and cosine, respectively. The
integral is indeed finite.

We therefore perform calculations for a set of values $\eta \in
(\eta_{min}, \eta_{max})$ for which $D(\eta)$ is a smooth
function with a well pronounced limit for large $R_{max}$. The
limit $\eta = 0$ is then determined at the end of calculations by
a quadratic least-square extrapolation method. Typically, 5-15
values of $\eta$ was used for $\eta_{min}\approx 0.5-0.6$ and
$\eta_{max}\approx 0.9-1.2$ with a relative error of order of a
few per cent. In calculations we have used $R_{max}$=7$a$ for fcc
and 9$a$ for bcc, where $a$ denotes the corresponding lattice
constant. It should be noted a proper order of limits in
Eq.~(\ref{eq_Dc}), namely first evaluate a sum for large
$R_{max}$ and then limit $\eta$ to zero. The procedure is
illustrated in Fig.~\ref{Fig.5}. The results for spin stiffness
coefficient $D$ calculated in this way are summarized in Table II
together with available experimental data \cite{ Mook, Pauth,
Shir}. There is a reasonable agreement between theory and
experiment for bcc-Fe and fcc-Co  but the values of spin-wave
stiffness constant are considerably overestimated for fcc-Ni. It
should be noted that measurements refer to the hcp-Co while the
present calculations were performed for fcc-Co. A similar
accuracy between calculated and measured spin-wave stiffness
constants was obtained by Halilov et al. \cite{ Halilov1998}
using the frozen-magnon approach. Our results are also in a good
agreement with those obtained by van~Schilfgaarde and Antropov
\cite{ Schilfgaarde1999} using the spin-spiral calculations to
overcome the problem of evaluation of $D$ from Eq.~(\ref{eq_D}).
On the other hand, this problem was overlooked in
Refs.~\onlinecite{ Liech, Spis, Antr2} so that a good agreement
of $D$, calculated for a small number of coordination shells,
with experimental data seems to be fortituous. Finally, results
of Brown et al. \cite{ Brown1999} obtained by the layer
Korringa-Kohn-Rostoker (KKR) method in the frozen potential
approximation are underestimated for all metals and the best
agreement is obtained for Ni.

\subsection{Curie temperature}

Several attempts have been made to evaluate Curie temperatures of magnetic
transition metals \cite{ Halilov1998, Schilfgaarde1999, Sakuma, Hubb, You} most of them based on the MFA.
The MFA as a rule overestimates values of Curie temperatures (with
exception of fcc-Ni with values substantially underestimated).
We will show that an alternative method based on the Green function
approach in the framework of the RPA \cite{ Tahi, Tahi2, Call, Tahi3}
can give a better agreement with experimental data.
The RPA Curie temperatures were calculated from Eq.~(\ref{eq_RPA1})
by employing the method of analytical deconvolution \cite{ deconv}.
In order to test the accuracy of this procedure we compare the
present numerical results for the ratio $T_C^{MFA}/T_C^{RPA}$
obtained for the nearest-neighbor Heisenberg model with the exact
results \cite{ Tahi, Tahi3}: we obtain 1.33 (fcc) and 1.37 (bcc)
as compared to exact values 1.34 and 1.39, respectively, i.e.,
a numerical procedure agrees with exact results within one per cent
accuracy.
Calculated values of Curie temperatures for both the MFA and RPA as well
as corresponding experimental data are summarized in Table II.
Mean-field values of Curie temperatures are overestimated for Fe and Co,
but underestimated for Ni in agreement with other calculations
\cite{ Halilov1998, Schilfgaarde1999}.
On the other hand, the results obtained using the RPA approach are
in a good agreement with experiment for both fcc-Co and bcc-Fe, while
the results for fcc-Ni are even more underestimated.
This is in agreement with the fact mentioned in Sec.~\ref{sec_Form},
namely that $T_C^{RPA} < T_C^{MFA}$.
The present results for Fe and Ni are in a good agreement with results
of Ref.~\cite{ Stau} using the spin-fluctuation theory and an improved
statistical treatment in the framework of the Onsager cavity-field
method.

The calculated ratio $T_C^{MFA}/T_C^{RPA}$ is 1.49, 1.25,
1.13 for bcc-Fe, fcc-Co, and fcc-Ni, respectively.
The values differ from those obtained for the first-nearest neighbor
Heisenberg model due to non-negligible next-nearest neighbors
in realistic ferromagnets and their oscillatory behavior with the shell
number.

The last point concerns the relevance of relativistic corrections
for the evaluation of the exchange parameters and related
quantities. The simplest quantity to evaluate is the MFA value of
the Curie temperature (see Eq.~(\ref{eq_J0})). Results for
ferromagnetic metals (including hcp-Co) are summarized in Table
III by comparing the non-relativistic and scalar-relativistic
values. One can conclude that scalar-relativistic corrections are
not important for fcc-Co and hcp-Co but their effect is
non-negligible for fcc-Ni and bcc-Fe. The scalar-relativistic
corrections generally shifts sp-bands downwards as compared to
the d-band complex while the changes of magnetic moments are
generally very small (a similar exchange splitting). One can thus
ascribe above changes mostly to the modifications of the density
of states at the Fermi energy (the site-diagonal blocks of the
Green function in Eq.~(\ref{eq_J0})). Results also show only a
weak dependence of the calculated $T_C^{MFA}$ on the structure
(hcp-Co vs fcc-Co).

\subsection{Comparison between the real-space and frozen
magnon approaches}

The real-space and frozen magnon approaches are formally
equivalent to each other. The quantities that are directly
calculated (the $J_{ij}$´s in the former case, the $E({\bf q})$'s
in the latter) are related to each other by a Fourier
transformation. Therefore, the pros and cons of both approaches
concern mainly their computational efficiency.

The computational effort needed to obtain one $J_{ij}$ parameter
within the real-space approach is approximately the same as to
compute one magnon energy $E({\bf q})$ within the frozen magnon
approach: in both cases a fine Brillouin zone integration is
required.

Therefore, it is quite clear that if one is primarily interested
in spin-wave dispersion curves (for a moderate number of ${\bf
q}$ points), or in the spin-wave stiffness $D$, the frozen magnon
approach is superior, for it does require to perform a Fourier
transformation and the delicate analysis explained in
Sec.~\ref{sec_sws}. We have shown, however, although less direct
and computationally more demanding, the real-space approach
performs well also.

On the other, if one is interested in the Curie temperature, the
real-space approach is more efficient. This obvious if ones uses
the mean-field approximation. Indeed, $T_C^{MFA}$ is obtained
from a {\em single\/} real-space calculation, by using the sum
rule (\ref{eq_sumrule}), whereas many $E({\bf q})$'s are needed
to obtain $T_C^{MFA}$ from Eq.~(\ref{eq_TcMFA}) within the frozen
magnon approach. Also if one uses the RPA, the real-space
approach is more efficient. For both approaches, the integral
over ${\bf q}$ in Eq.~(\ref{eq_TcRPA}) needs to be performed
accurately, with paying great attention to the divergence of the
integrant at ${\bf q}=0$. A very high density of ${\bf q}$ points
is required there, in order to have a satisfactory convergence.
Within the frozen-magnon approach, each of the $E({\bf q})$'s
requires the same computational effort. In contrast, within the
real-space approach, less than 200 $J_{ij}$'s are sufficient to
obtain a parametrization of $E({\bf q})$ over the full Brillouin
zone, which considerably reduces the computational effort.

A further very important advantage of the real-space approach is
its straightforward application to systems with broken
translational symmetry like, e.g., substitutional alloys,
surfaces, overlayers, and multilayers. This is an important
advantage keeping in mind the relevance and yet not fully
understood character of exchange interactions at metal interfaces
and surfaces. The reciprocal-space approach can be applied to
ideal surfaces, but it is numericallly demanding, and its
application to system with substitutional disorder and/or to
finite magnetic clusters is practically impossible. Finally, the
dependence of exchange parameters $J_{ij}$ on the distance also
gives an important information about the nature of the magnetic
state (RKKY-like interactions) and this dependence is again
straigtforwardly determined by the real-space method while in the
reciproacal-space method $J_{ij}$'s have to be determined by
inverting Eq.~(\ref{eq_E}).

\section{Summary}
We have calculated Heisenberg exchange parameters of bcc-Fe,
fcc-Co, and fcc-Ni in real space from first-principles by
employing the magnetic force theorem. We have determined
dispersion curves of magnetic excitations along high-symmetry
directions in the Brillouin zone, spin-wave stiffness constants,
and Curie temperatures of considered metals on the same footing,
namely all based on calculated values of exchange parameters
$J_{ij}$. Dispersion curves of bcc Fe exhibit an anisotropic
behavior in the range of long wavelengths, with peculiar minima
for short wavelengths in the [100]-direction which are due to a
relatively strong exchange oscillations in this metal. We have
presented a method of evaluation of the spin-wave stiffness
constants which yields converged values, in contrast to previous
results in the literature. Calculated spin-wave stiffness
constants agree reasonably well with available experimental data
for Co and Fe, while agreement is rather poor for Ni. Present
calculations agree also well with available experimental data for
magnon dispersion law of bcc-Fe. We have also evaluated Curie
temperatures of metals in question using the mean-field
approximation and the Green function random phase approximation.
We have found that in the latter case a good agreement with the
experiment is obtained for Co and Fe, while less satisfactory
results are obtained for Ni, where the role of the Stoner
excitations is much more important as compared to Co and Fe. In
addition, the adiabatic approximation is less justified for Ni,
and, possibly, correlation effects beyond the local density
approximation play the more important role for this ferromagnet.

In conclusion, we have demonstrated that the real-space appraoch
is able to determine the low-lying excitations in ferromagnetic
metals with an accuracy comparable to the reciprocal-space
approach. This justifies the use of the real-space approach for
more interesting and complex systems with violated translational
symmetry where the reciprocal-space approach is of the limited
use. In particular, a first promising application of the
real-space approach to the problem of the oscillatory Curie
temperature of two-dimensional ferromagnets has been recently
published \cite{ oscTc}.

\section{Acknowledgments}

J.K., V.D., and I.T. acknowledge financial support provided by the
Grant Agency of the Academy of Sciences of the Czech Republic
(Project A1010829), the Grant Agency of the Czech Republic
(Project 202/00/0122), and the Czech Ministry of Education, Youth,
and Sports (Project OC P3.40 and OC P3.70).

\newpage

\begin{table}
\caption{Effective Heisenberg exchange parameters $J_{0j}$ for ferromagnetic Fe, Co,
and Ni for the first 10 shells. Quantities {\bf R}$_{0j}$ and $N_s$ denote,
respectively, shell coordinates in units of corresponding lattice constants and the
number of equivalent sites in the shell.}
\begin{center}
\begin{tabular}{crdcrdcrd}
  & Fe (bcc) &  &  & Co (fcc) &  &  &  Ni (fcc) &  \\ \tableline
  {\bf R}$_{0j}$ & $N_s$ & $J_{0j}$[mRy] & {\bf R}$_{0j}$ & $N_s$ & $J_{0j}$[mRy] & {\bf
 R}$_{0j}$ & $N_s$ & $J_{0j}$[mRy] \\ \tableline
($\frac{1}{2} \frac{1}{2} \frac{1}{2}$) &  8 & 1.432 & ($\frac{1}{2} \frac{1}{2} 0$) &
12
 & 1.085 & ($\frac{1}{2} \frac{1}{2} 0$) & 12 & 0.206 \\
($1 0 0$) & 6 &  0.815 & ($1 0 0$) & 6 & 0.110 & ($1 0 0$) & 6 &  0.006 \\
($1 1 0$) & 12 & $-$0.016 & ($1 \frac{1}{2} \frac{1}{2}$) & 24 &  0.116 & ($1 \frac{1}{2
} \frac{1}{2}$) & 24 &  0.026 \\
($\frac{3}{2} \frac{1}{2} \frac{1}{2}$) & 24 & -0.126 & ($1 1 0$) & 12 & $-$0.090 & ($1
1 0$) & 12 &  0.012 \\
($1 1 1$) & 8 & -0.146 & ($\frac{3}{2} \frac{1}{2} 0$) & 24 &  0.026 & ($\frac{3}{2}
\frac{1}{2} 0$) & 24 &  0.003 \\
($2 0 0$) & 6 &  0.062 & ($1 1 1$) & 8 &  0.043 & ($1 1 1$) & 8 & $-$0.003 \\
($\frac{3}{2} \frac{3}{2} \frac{1}{2}$) & 24 &  0.001 & ($\frac{3}{2} 1 \frac{1}{2}$) &
48 & $-$0.024 & ($\frac{3}{2} 1 \frac{1}{2}$) & 48 &  0.007 \\
($2 1 0$) & 24 &  0.015 & ($2 0 0$) & 6 &  0.012 & ($2 0 0$) & 6 & $-$0.001 \\
($2 1 1$) & 24 & $-$0.032 & ($\frac{3}{2} \frac{3}{2} 0$) & 12 &  0.026 & ($\frac{3}{2}
\frac{3}{2} 0$) & 12 & $-$0.011 \\
($\frac{3}{2} \frac{3}{2} \frac{3}{2}$) & 8 &  0.187 & ($2 \frac{1}{2} \frac{1}{2}$) & 2
4 &  0.006 & ($2 \frac{1}{2} \frac{1}{2}$) & 24 &  0.001 \\
\end{tabular}
\end{center}
\end{table}

\begin{table}[h]
\caption{ Calculated spin-wave stiffness constants ($D_{th}$) and Curie
 temperatures
($T_C^{MFA}$ and $T_C^{RPA}$) and their comparison with experimental values
 $D_{ex}$ and $T_C^{ex}$.}
\begin{center}
\begin{tabular}{clrrrr}
Metal & $D_{th}$[meV$\cdot {\rm \AA}^2$] & $D_{ex}$[meV$\cdot {\rm \AA}^2$] &
$T_C^{MFA}$[K]
 & $T_C^{RPA}$[K] & $T_C^{ex}$[K]\\ \hline\hline
Fe (bcc) & $250\pm7$ & $280^c,330^d$ & $1414$ & $950\pm2$ & $1044-1045$\\
Co (fcc) & $663\pm6$ & $580^{a,c}$,$510^d$ & $1645$ & $1311\pm4$ & $1388-1398^a$\\
Ni (fcc) & $756\pm29$ & $555^b,422^c$ & $397$ & $350\pm2$ & $624-631$\\
\end{tabular}
\end{center}
$^a$Data refer to hcp Co at 4.2 K.\\
$^b$Neutron scattering measurement at 4.2 K \cite{ Mook}. \\
$^c$Magnetization measurement \cite{ Pauth} at 4.2 K.\\
$^d$Neutron scattering measurement extrapolated to 0 K \cite{ Shir}. \\
\end{table}

\newpage

\begin{table}[h]
\caption{ Calculated Curie temperatures of ferromagnetic metals in the
mean-field approximation for non-relativistic ($nr$) and
scalar-relativistic ($sr$) cases.}
\begin{center}
\begin{tabular}{rrr}
Metal & $T_C^{nr}$[K]  & $T_C^{sr}$[K] \\ \hline\hline
Fe (bcc) & $1414$ & $1335$ \\
Co (fcc) & $1645$ & $1651$ \\
Co (hcp) & $1679$ & $1673$ \\
Ni (fcc) & $ 397$ & $ 428$ \\
\end{tabular}
\end{center}
\end{table}
\newpage

\begin{figure}
\caption{Magnon dispersion relations along high-symmetry lines in the
 Brillouin zone: (a) bcc-Fe (experiment: Ref.~[33], 10 K, filled circles
 and Ref.~[35], Fe(12 \% Si), room temperature, empty squares); (b) fcc-Co;
 and (c) fcc-Ni (experiment: Ref.~[34], room temperature, empty circles).
 Lines are calculated results.
}
\label{Fig.1}
\end{figure}


\begin{figure}
\caption{Spin-wave stiffness constants calculated from Eq.~(\ref{eq_Dc})
as a function of $R_{max}$ (in units of lattice constants) for fcc-Ni
(full line), fcc-Co (short dashes), and bcc-Fe (long dashes).}
\label{Fig.3}
\end{figure}

\begin{figure}
\caption{Spin-wave stiffness of fcc-Ni calculated from Eq.~(\ref{eq_D})
as a function of $R_{max}$ (in units of lattice constant) for various
values of the damping factor $\eta$.}
\label{Fig.4}
\end{figure}

\begin{figure}
\caption{Spin-wave stiffness coefficients $D(\eta)$ for bcc-Fe
(empty squares), fcc-Co (empty triangles), and fcc-Ni (empty circles) as
a function of the parameter $\eta$ and extrapolated values for
$\eta=0$ (filled symbols). The solid line indicates the quadratic fit
function used for extrapolation.}
\label{Fig.5}
\end{figure}

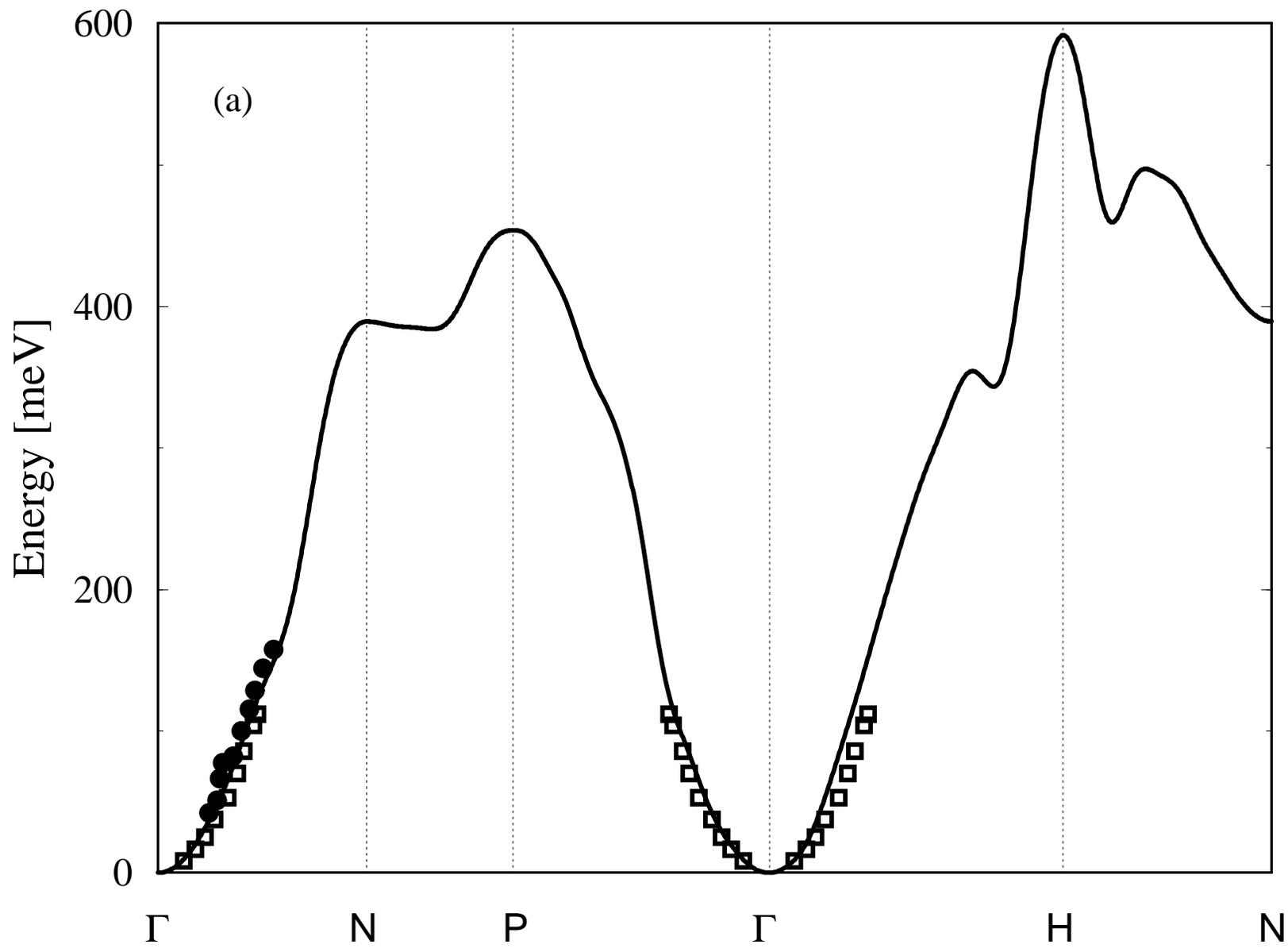
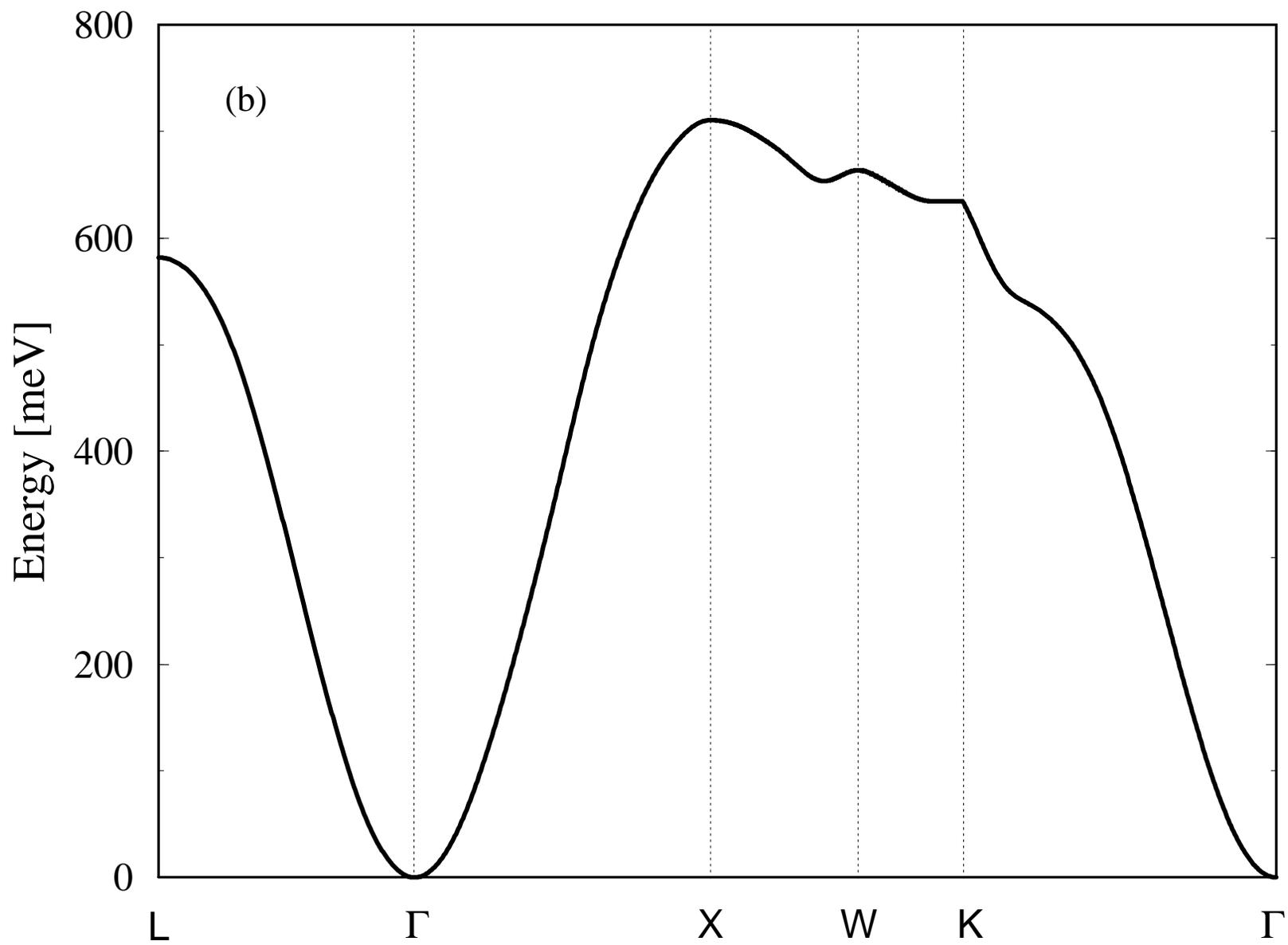
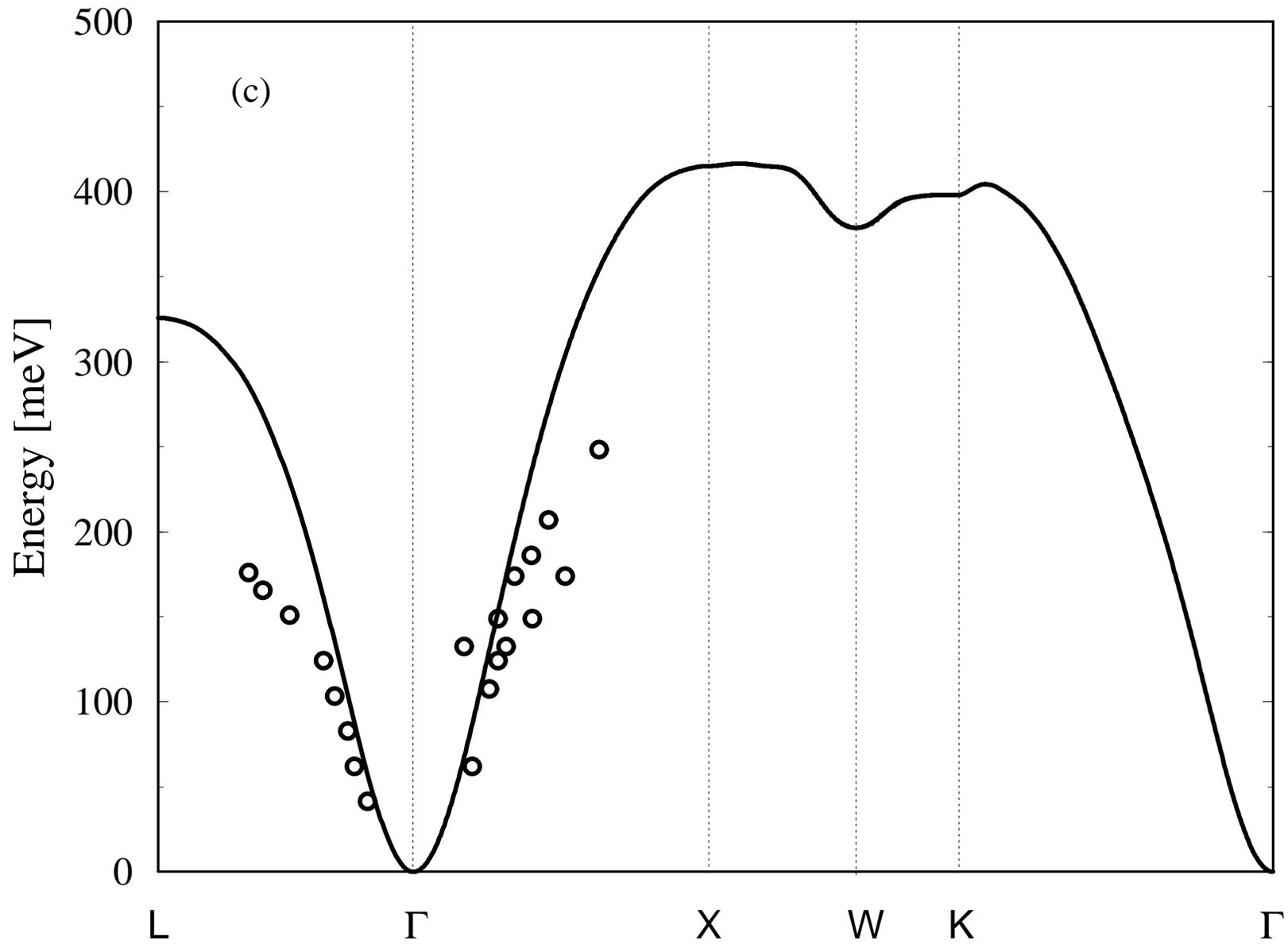
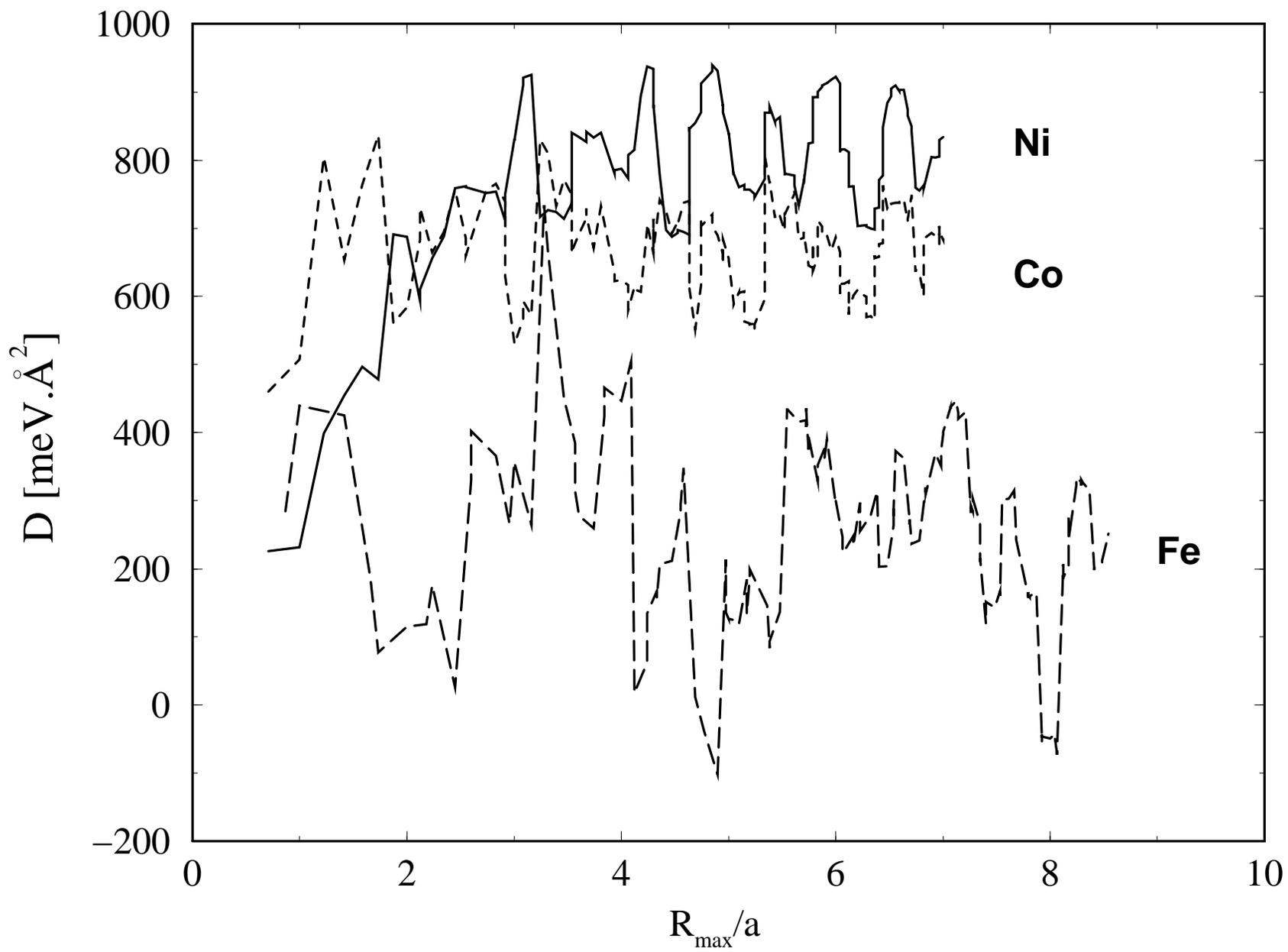
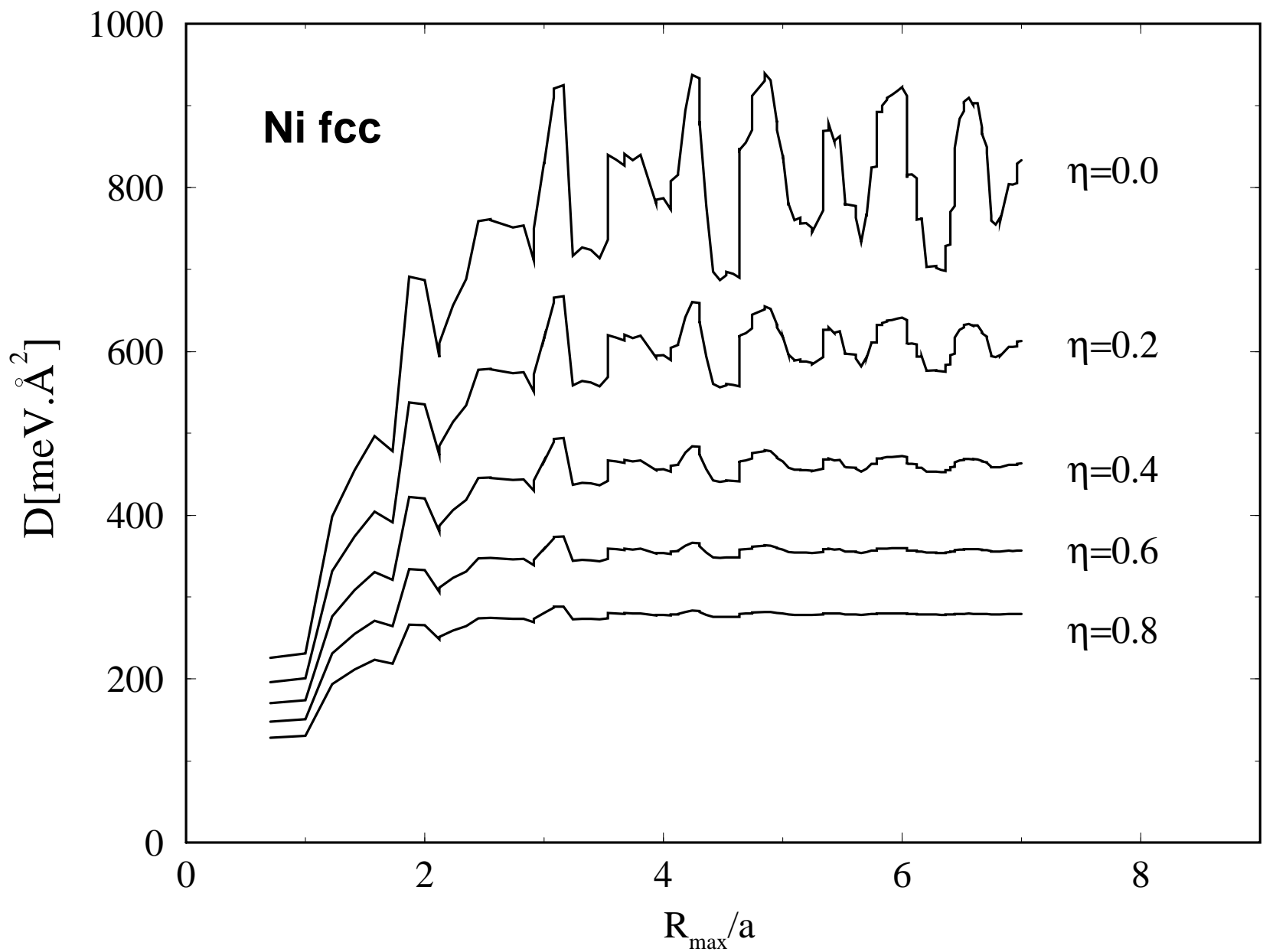
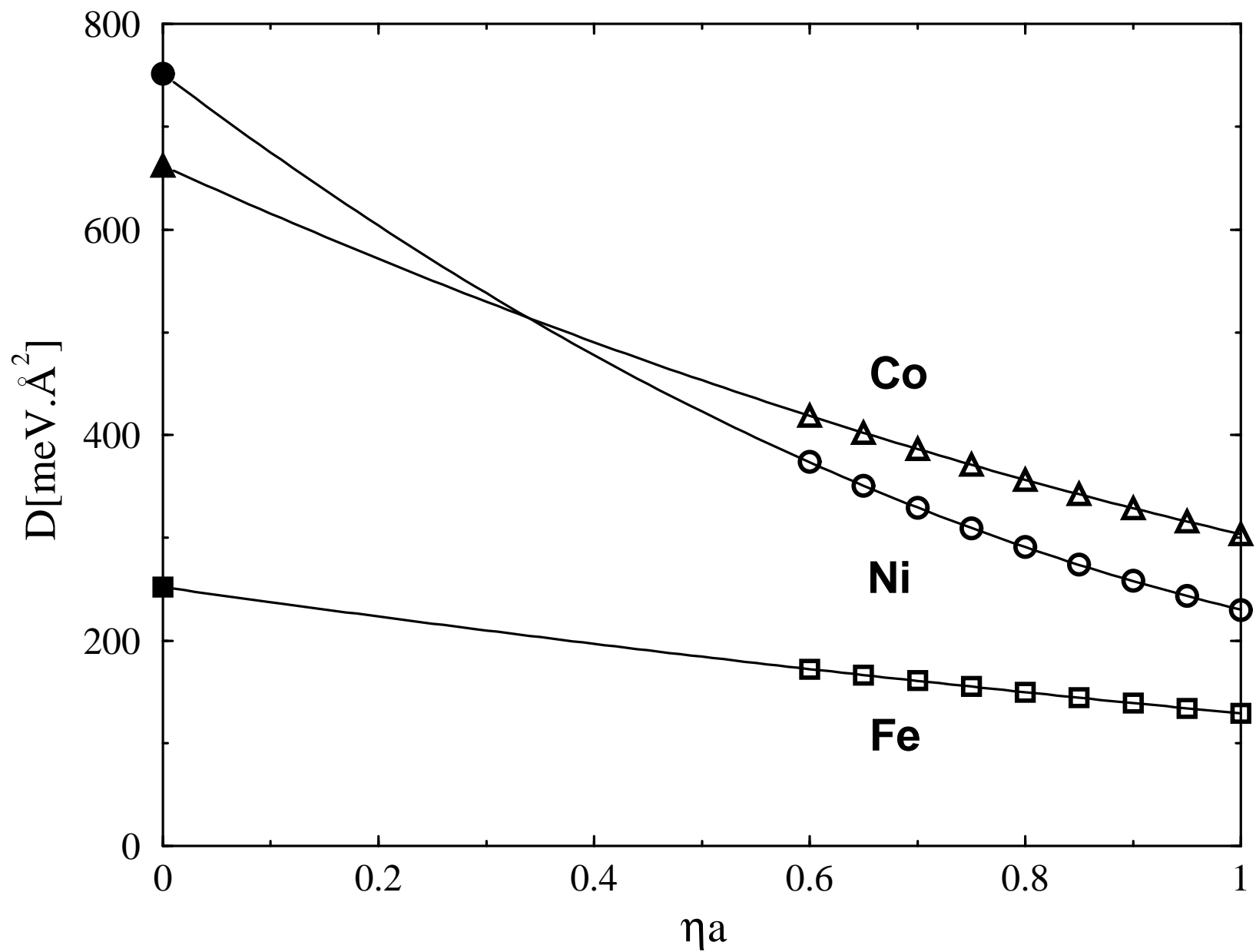
\end{document}